\documentclass[%
 aip,
 amsmath,amssymb,
 reprint,%
]{revtex4-2}

\usepackage{dcolumn}
\usepackage{bm}
\usepackage{braket}
\usepackage{color}
\usepackage{natbib}
\usepackage[normalem]{ulem}
\usepackage{ifpdf}
\ifpdf
  \usepackage{graphicx}
\else
  \usepackage[dvipdfmx]{graphicx}
\fi

\begin{document}


\title{
Cryogenic Time-Division-Multiplexed Voltage Control for Scalable Trapped-Ion Quantum Processors
}

\author{Ryutaro~Ohira}
\email{
ohira@quel-inc.com
}
\affiliation{
QuEL, Inc., Hachioji ON Building 5F, 4-7-14 Myojincho, Hachioji, Tokyo, Japan
}

\author{Shinichi~Morisaka}
\affiliation{
QuEL, Inc., Hachioji ON Building 5F, 4-7-14 Myojincho, Hachioji, Tokyo, Japan
}
\affiliation{
Center for Quantum Information and Quantum Biology, The University of Osaka, 1-2 Machikaneyama, Toyonaka, Osaka, Japan
}

\author{Yoshinori~Kurimoto}
\affiliation{
QuEL, Inc., Hachioji ON Building 5F, 4-7-14 Myojincho, Hachioji, Tokyo, Japan
}

\author{Toshiaki~Inada}
\affiliation{
International Center for Elementary Particle Physics (ICEPP), The University of Tokyo, 7-3-1 Hongo, Bunkyo, 113-0033, Tokyo, Japan
}

\author{Ippei~Nakamura}
\affiliation{
Komaba Institute for Science (KIS), The University of Tokyo, 3-8-1 Komaba, Meguro, 153-8902, Tokyo, Japan
}

\author{Takefumi~Miyoshi}
\affiliation{
QuEL, Inc., Hachioji ON Building 5F, 4-7-14 Myojincho, Hachioji, Tokyo, Japan
}
\affiliation{
Center for Quantum Information and Quantum Biology, The University of Osaka, 1-2 Machikaneyama, Toyonaka, Osaka, Japan
}
\affiliation{
e-trees. Japan, Inc., Daiwaunyu Building 2F, 2-9-2 Owadamachi, Hachioji, Tokyo, Japan
}

\author{Atsushi~Noguchi}
\affiliation{
Komaba Institute for Science (KIS), The University of Tokyo, 3-8-1 Komaba, Meguro, 153-8902, Tokyo, Japan
}
\affiliation{
RIKEN Center for Quantum Computing (RQC), 2-1 Hirosawa, Wako, 351-0198, Saitama, Japan
}
\affiliation{
Inamori Research Institute for Science (InaRIS), 620 Suiginya, Kyoto, 600-8411, Kyoto, Japan
}

\date{\today}

\begin{abstract}
Trapped-ion quantum computers based on the quantum charge-coupled device architecture require on the order of ten trap electrodes per qubit, making the number of vacuum feedthroughs a bottleneck at the system scale.
Time-division multiplexed (TDM)-based voltage control for trap electrodes provides a natural route to alleviate this constraint. 
However, previous studies have been limited to architectural proposals for static trap-potential compensation and room-temperature demonstrations of dynamic-electrode control, leaving cryogenic operation of TDM-based voltage control for static and dynamic electrodes experimentally unexplored.
In this study, we develop and cryogenically validate TDM-based voltage control schemes for two distinct electrode classes. 
For static electrodes used in trap-potential compensation, we implement a 32-channel demultiplexed system operating at approximately 27~K, achieving an effective voltage update rate of 37.5~kHz with an output range of $\pm10~\mathrm{V}$ per channel. 
For dynamic electrodes used in ion operations, such as shuttling, we implement a four-channel demultiplexed system operating at approximately 14~K, achieving an effective voltage update rate of 1~MHz with a comparable output range.
These results establish TDM-based voltage control as a practical approach for both electrode classes, providing a path for mitigating the vacuum feedthrough bottleneck in scalable trapped-ion quantum processors.
\end{abstract}

\maketitle


Quantum charge-coupled device (QCCD) architectures are among the leading platforms for scalable quantum computing with trapped ions, where trapped-ion qubits are dynamically shuttled among dedicated zones for state preparation, gate operations, and measurement~\cite{wineland1998experimental, kielpinski2002architecture}. 
Recent experimental progress in this area has been remarkable, with leading platforms now approaching the 100-qubit regime~\cite{ransford2025helios}. 
A central challenge in scaling these systems is the large electrode count required: approximately ten trap electrodes per qubit~\cite{malinowski2023wire}.
Consequently, the current 98-qubit state-of-the-art system already employs more than one thousand electrodes~\cite{ransford2025helios}. 
This rapid growth in the number of electrode places severe pressure on the limited number of available vacuum feedthroughs, creating a wiring bottleneck that becomes increasingly difficult to address as the system scales~\cite{malinowski2023wire}.

Several strategies have been developed to mitigate this bottleneck. 
Voltage broadcasting shares a common signal source among multiple electrodes and is well-suited to conveyor-belt-style ion transport; however, adapting to more complex dynamic operations that require independent electrode control is difficult~\cite{moses2023race, ransford2025helios, delaney2024scalable}.
Cryogenic electronics reduce wiring requirements by integrating active components inside the cryogenic stage within the vacuum chamber; however, the number of required electronic components still scales with the number of electrodes~\cite{stuart2019chip, sieberer2024cryogenic, park2024cryo, meyer2025development, meyer2025cryogenic}. 
Moreover, the voltages of $\pm 10~\mathrm{V}$ required for trapped-ion control necessitate amplifiers, whose heat dissipation places significant demands on the limited cooling capacity of cryogenic systems.

Time-division multiplexed (TDM)-based voltage control provides a structurally different solution~\cite{hunter2024methods, malinowski2023wire, jones2025architecting}. 
TDM routes multiple electrode voltages through a single control line and demultiplexes them locally near the trap, allowing a single feedthrough to serve $N$ electrodes, reducing the number of required feedthroughs compared with a conventional one-line-per-electrode architecture. 
TDM-based voltage control was first explored in the context of the wiring using integrated switching electronics (WISE) architecture~\cite{malinowski2023wire}, which classifies electrodes into two functional classes: quasistatic electrodes, whose voltages are set and held during ion operations, and dynamic electrodes, which must be driven by time-dependent waveforms for shuttling, splitting, merging, and rotation. 
However, previous research remains an architectural proposal and has not been experimentally demonstrated. 
Moreover, the TDM-based voltage control scheme proposed in the WISE architecture is primarily designed for quasistatic electrodes; the dynamic electrode class, which requires MHz-level voltage update rates, remains a more challenging target and has not been addressed experimentally.

Our recent work introduced and experimentally demonstrated a TDM-based voltage control framework for dynamic electrodes~\cite{ohira2025multiplexed, ohira2025trapping}, establishing its feasibility for ion trapping and transport. 
However, these demonstrations were conducted at room temperature with the demultiplexer placed outside the vacuum chamber. 
Therefore, two critical questions remain unanswered: (1)~Can TDM-based voltage control be implemented cryogenically, where ion-trap systems operate and the wiring reduction benefit is realized? (2)~Does TDM-based voltage control meet the performance requirements of both electrode classes, which are the modest update rates required for static compensation and the MHz-level rates required for dynamic operations?

This study answers both questions by developing a TDM-based voltage control system for cryogenic operation and characterizing two prototype systems covering the full range of QCCD electrode requirements.
The first system is a 32-channel demultiplexed system for static-electrode control, operated at approximately 27~K, achieving an effective voltage update rate of 37.5~kHz with $\pm10$~V per channel. 
The second system is a four-channel system for dynamic-electrode control, operated at approximately 14~K, achieving an update rate of 1~MHz over a comparable voltage range. 
Collectively, these results establish cryogenic TDM-based voltage control as a scalable solution to the wiring bottleneck in trapped-ion quantum computing.

\begin{figure}
    \centering
    \includegraphics[width=8.5cm]{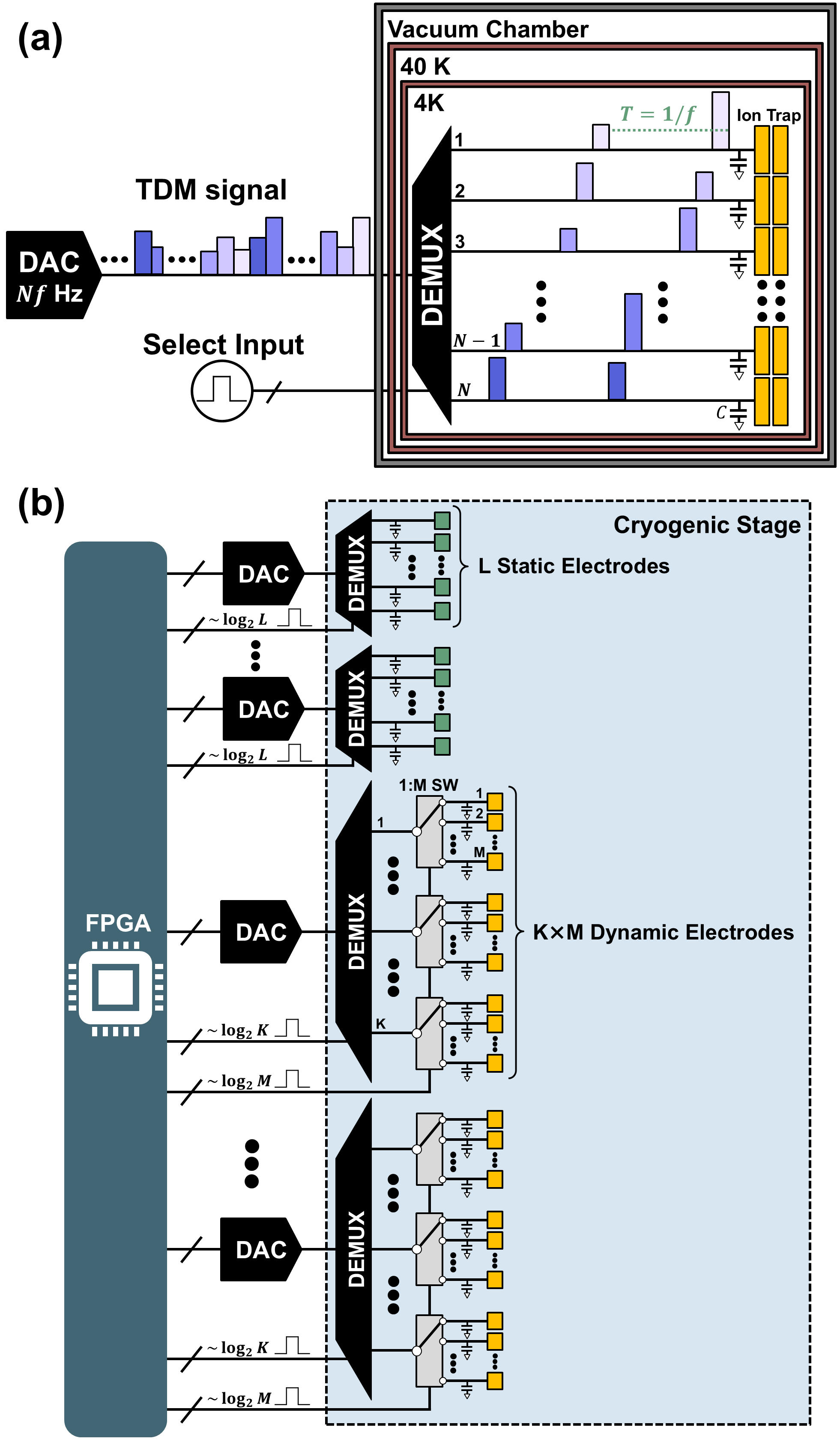}
    \caption{
    (a) Schematic of a TDM-based voltage control system for cryogenic ion traps. 
    A representative cryogenic setup is shown, where a vacuum chamber contains 40-K and 4-K stages and the ion trap is mounted on the 4-K stage. 
    A DAC at room temperature generates a TDM waveform at an update rate of $Nf~\mathrm{Hz}$, where $N$ represents the number of trap electrodes. 
    The waveform is sent to a demultiplexer (denoted as DEMUX) located at the 4-K stage, which distributes the signal to the individual electrodes. 
    A holding capacitor connected to each demultiplexer output stores the voltage for the corresponding electrode via charging during the selected time slot. 
    Consequently, each electrode is updated at an effective rate of $f~\mathrm{Hz}$. 
    Select Input denotes the digital control signal for the demultiplexer. 
    (b) Schematic of TDM-based voltage control architecture for static and dynamic trap electrodes. 
    For dynamic-electrode control, because not all electrodes require time-varying waveforms simultaneously, an additional switch network after the demultiplexer output enables selective routing of the TDM signal to the active electrode group, thereby extending effective control to $K \times M$ dynamic electrodes.
    }
    \label{Fig:archi}
\end{figure}
%


Here, we describe the TDM-based voltage control architecture proposed in our previous work~\cite{ohira2025multiplexed,ohira2025trapping}.
Figure~\ref{Fig:archi}(a) illustrates the overall concept.
In this study, we consider a cryogenic ion-trap system based on a typical cryogenic-stage configuration with 40-K and 4-K stages~\cite{pagano2019cryogenic}.
At room temperature, a single digital-to-analog converter (DAC) generates a TDM waveform at an update rate of $Nf~\mathrm{Hz}$ and sequentially outputs the target voltages for all $N$ electrodes within each multiplexing period.
The waveform is transmitted into the cryostat and delivered to a demultiplexer located at the 4-K stage, which routes each voltage sample to the corresponding electrode.
A holding capacitor at each output retains the assigned voltage until the next update cycle.

In this setup, synchronization between the DAC output and the digital control signal for the demultiplexer (denoted as Select Input in Fig.~\ref{Fig:archi}(a)) is critical: each capacitor must be fully charged to its target voltage within the allotted time slot of duration $1/Nf$. 
As analyzed in Ref.~\cite{ohira2025multiplexed}, this window must accommodate both the DAC output settling and the subsequent capacitor charging, imposing a constraint on the RC time constant of each output node. 
Detailed analysis of this constraint and the resulting design requirements is provided in Ref.~\cite{ohira2025multiplexed}.

In this study, we consider a surface-electrode ion-trap layout following the WISE architecture discussed by Malinowski et al.~\cite{malinowski2023wire}.
In this layout, many DC control electrodes are distributed along the trap axis and can be grouped into two principal classes: static electrodes used for trap-potential compensation and dynamic electrodes used for shuttling, splitting, merging, and rotation.
Based on this classification, we describe how our TDM-based voltage-control architecture can be applied to each electrode class in QCCD systems.
The basic configuration is common to both static- and dynamic-electrode control.
At room temperature, a digital controller, such as a field-programmable gate array (FPGA), drives a DAC to generate the time-division-multiplexed voltage waveform required for the target electrodes.
The same digital controller is also assumed to generate the logic signals used to control the demultiplexer.
The multiplexed analog and demultiplexer control signals are then routed to the cryogenic stage, where the demultiplexer distributes the analog voltage to the selected electrode outputs.

Static electrodes compensate for trapping potential deviations arising from fabrication imperfections and environmental disturbances. 
They require stable DC voltages, typically within a range of $\pm 10~\mathrm{V}$, and are updated infrequently compared with the timescale of ion operations. 
In the proposed scheme, the target DC voltages for all static electrodes are time-division multiplexed at room temperature and distributed by a cryogenic demultiplexer, as illustrated in Fig.~\ref{Fig:archi}(b).

Dynamic electrodes execute ion-transport operations and must therefore deliver time-varying arbitrary waveforms over a voltage range of $\pm 10~\mathrm{V}$, with per-electrode update rates on the order of megahertz.
Although the basic TDM configuration for dynamic electrodes is the same as that for static electrodes, dynamic-electrode control has an additional scaling feature.
In a typical QCCD operation, only the electrodes in the active transport zone require time-varying waveforms at a given moment, whereas electrodes in idle zones do not require dynamic updating.
This allows an additional switch network to be introduced after the demultiplexer outputs, enabling selective routing of the TDM signal to the currently active electrode group.
As shown in Fig.~\ref{Fig:archi}(b), the proposed architecture extends effective control to $K \times M$ dynamic electrodes while maintaining the same number of external connections.

Although the number of available FPGA pins is generally limited, the system can be further scaled by operating multiple TDM-based voltage-control units, such as the one shown in Fig.~\ref{Fig:archi}(b), in parallel with high-precision synchronization.
One possible synchronization approach is Gigabit-Ethernet-based coordination among FPGA units~\cite{miyoshi2025toward, kurimoto2026microwave}.


%
\begin{figure}[t]
    \centering
    \includegraphics[width=8.5cm]{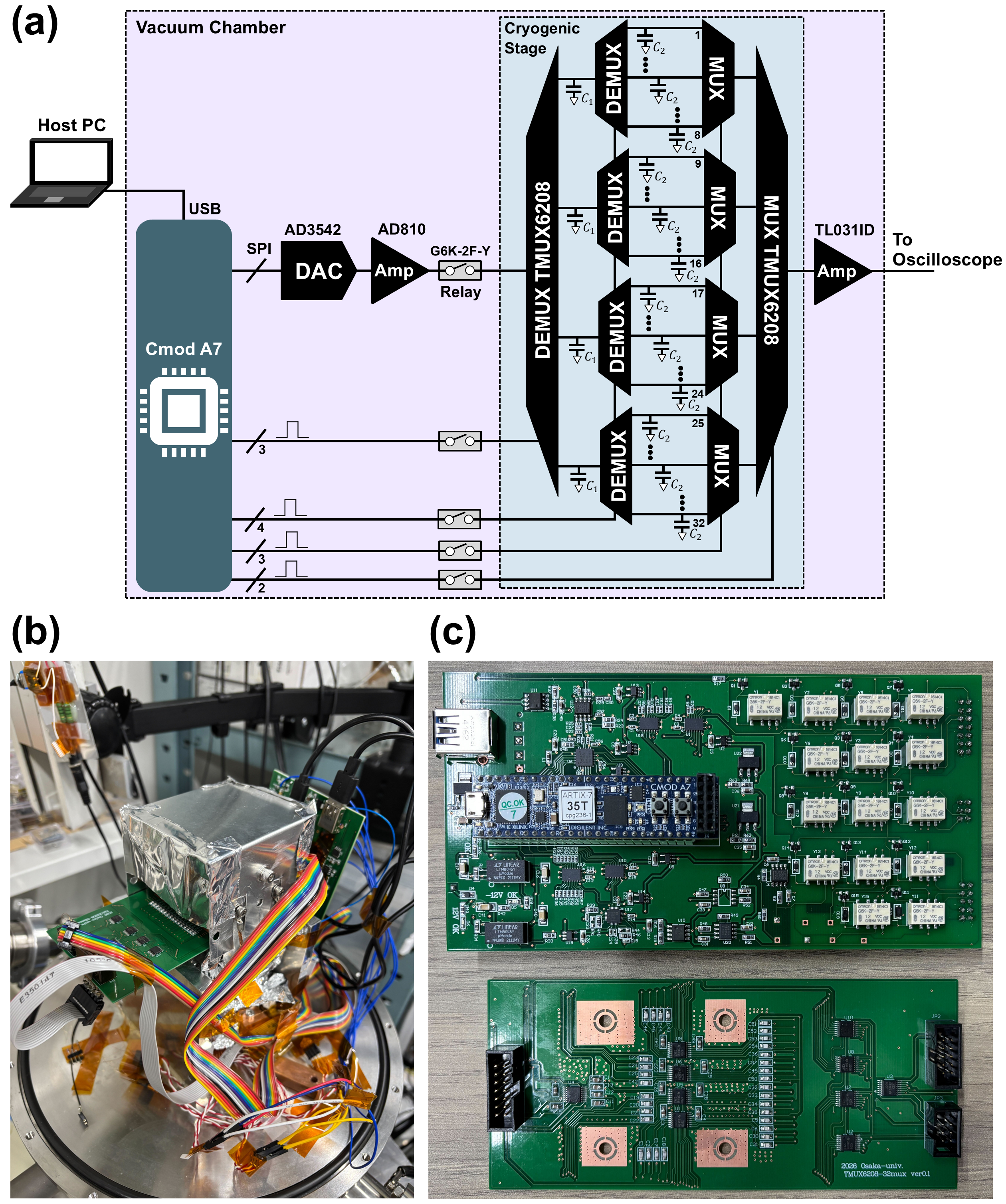}
    \caption{
    TDM-based voltage control system for static electrodes.
    (a) Schematic of the system architecture.
    (b) Cryogenic measurement setup with the vacuum chamber removed.
    (c) TDM signal generation board and DEMUX board.
    The TDM signal generation board, shown at the top, includes an FPGA, a DAC, operational amplifiers, and relay switches.
    The 32-channel DEMUX board, shown at the bottom, is installed on the cryogenic stage.
    }
    \label{Fig:static_control_setup}
\end{figure}

Based on the proposed architecture, we develop prototype TDM-based voltage control systems for both static and dynamic electrodes and validate their operation in cryogenic environments.
First, we describe the TDM-based voltage control system developed for static-electrode control and present a cryogenic proof-of-concept evaluation of its operation.

For static-electrode control, we develop a 32-channel multiplexed voltage control system.
A schematic of the overall system is shown in Fig.~\ref{Fig:static_control_setup}(a), and a photograph of the cryogenic measurement setup is shown in Fig.~\ref{Fig:static_control_setup}(b).

The system comprises two main boards: (1) a TDM signal generation board and (2) a DEMUX board.
The TDM signal generation board, shown in the upper part of Fig.~\ref{Fig:static_control_setup}(c), generates multiplexed voltage signals at room temperature. 
This board includes an FPGA, a DAC, operational amplifiers, and relay switches.
The TDM signal generation board is physically placed inside the vacuum chamber while remaining at room temperature, and the FPGA on the board is controlled through the host PC through a single USB connection. 
TDM signals are generated by a DAC (AD3542, Analog Devices, Inc.) controlled via SPI by a Digilent Cmod A7 FPGA board. 
The DAC operates at an update rate of 1.5~MHz, and its output is amplified to approximately $\pm10~\mathrm{V}$ by an operational amplifier (AD810, Analog Devices, Inc.). 
Each output channel is equipped with a relay that only transmits the signal during operation.

The amplified TDM signal is transmitted to a DEMUX board, which is shown in lower part of Fig.~\ref{Fig:static_control_setup}(c).
This board is installed on the cryogenic stage, where the signal is distributed to individual output channels.
The DEMUX board is implemented using a two-stage configuration of 8:1 multiplexers (TMUX6208, Texas Instruments Inc.), with four and eight outputs at the first and second stages, respectively, yielding up to $4\times8=32$ channels. 
Although the configuration can, in principle, support up to $8\times 8 = 64$ channels, the current implementation is limited to 32 channels by the available installation area on the cryogenic stage, which is approximately $70~\mathrm{mm} \times 50~\mathrm{mm}$.

As shown in Fig.~\ref{Fig:static_control_setup}(a), the DEMUX board includes holding capacitors $C_1$ (2.7~nF) and $C_2$ (470~pF) connected at the first and second stages, respectively. 
In principle, a multistage demultiplexer only requires a holding capacitor at the final output stage. 
However, TMUX6208 has an on-capacitance of approximately 185~pF~\cite{TI_TMUX6208_datasheet}, which causes voltage division at the first stage and prevents the second-stage capacitor from charging to the target voltage within the available time slot. 
This issue can be addressed by introducing $C_1$ at the first stage and adopting a two-step charging sequence: the four first-stage capacitors are charged sequentially before the corresponding second-stage switch is activated. 
This sequence requires five time slots to refresh four channels, yielding an efficiency factor of 4/5. 
Because this is repeated across all eight second-stage outputs, one full refresh of 32 channels requires 40 DAC update steps, yielding an effective per-channel update rate of $1.5~\mathrm{MHz}/40=37.5~\mathrm{kHz}$, compared with the nominal $1.5~\mathrm{MHz}/32\approx46.9~\mathrm{kHz}$.

Another important feature of the system is the readout configuration used for cryogenic evaluation.
The board provides 32 output channels; however, directly monitoring all of them would require 32 separate lines from the cryogenic stage to room temperature, substantially complicating the measurement setup.
To prevent this complication, the 32 output channels are multiplexed again using additional 8:1 multiplexers (TMUX6208), allowing all outputs to be monitored through a single readout line.
This signal is returned to the TDM signal generation board, buffered using an operational amplifier (TL031ID, Texas Instruments Inc.), and then measured at room temperature using an oscilloscope, as shown in Fig.~\ref{Fig:static_control_setup}(a).

For the cryogenic evaluation, we use an RDC-02K 2K cryocooler (Sumitomo Heavy Industries, Ltd.).
As shown in Fig.~\ref{Fig:static_control_setup}(b), the TDM signal generation board is installed inside the vacuum chamber and positioned near the DEMUX board.
Because the FPGA on the TDM signal generation board generates non-negligible heat inside the vacuum chamber, the DEMUX board is thermally shielded, and the ribbon cables connecting the two boards are thermally anchored at the 40-K stage.
With this configuration, the system reaches a stable temperature of approximately 27~K after cooling.

The relatively high operating temperature is primarily attributed to the limited cooling power of the cryocooler used in this setup.
The specified cooling power is 1.0~W at 60~K for the first stage and 0.02~W at 2.3~K for the second stage~\cite{RDC_02K}.
In addition, because the room-temperature TDM signal generation board is located in proximity to the 4-K stage, thermal radiation from the board is likely to provide an additional heat load.
Furthermore, the temperature sensor is mounted on the top surface of the printed circuit board under evaluation; therefore, the measured temperature may be higher than the temperature of the cold head.
Nevertheless, in a separate experiment using another cryocooler, we confirm that the TMUX6208 switch used in this study operates at approximately 2~K with no noticeable degradation in its basic switching behavior.


%
\begin{figure}[t]
    \centering
    \includegraphics[width=8.5cm]{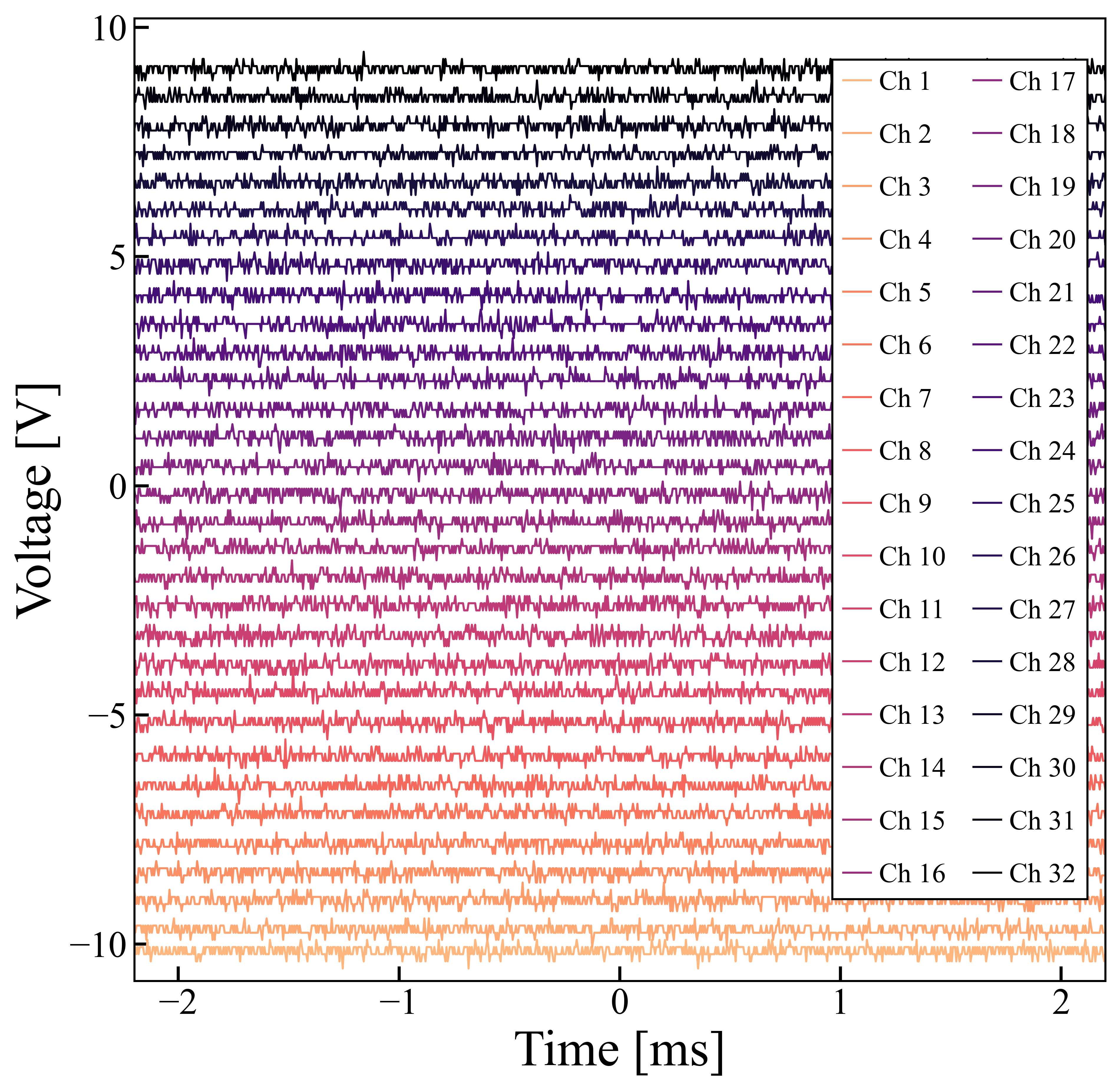}
    \caption{
    Measured DC output voltages of the 32-channel multiplexed system at approximately 27~K.
    }
    \label{Fig:static_control_result}
\end{figure}

To verify the basic operation of the proposed system, an experiment is conducted in which constant-voltage signals ranging from $-10$ to $+10~\mathrm{V}$ are multiplexed in time, demultiplexed using the cryogenic circuit, and measured at the output.
Figure~\ref{Fig:static_control_result} shows the measured DC output voltages of the 32-channel multiplexed system at approximately 27~K.
The results confirm that the developed TDM-based voltage control system successfully distributes the programmed voltage values to all 32 output channels.
Each channel exhibits the intended DC output level, indicating that both the demultiplexing operation and the voltage-holding mechanism function adequately in the cryogenic environment.

Voltage fluctuations are observed in the measured output waveforms.
These fluctuations can be attributed to charge injection into the ground when the demultiplexer (TMUX6208) switches to the ON state. 
In addition, the 32 output levels are not equally spaced, indicating interchannel crosstalk.
We evaluate this effect at cryogenic temperature by applying a large-amplitude sinusoidal waveform to channel 1 and measuring the induced signals on the other channels. 
In the worst case, the induced signal on channel 2 corresponds to a crosstalk level of approximately -12~dB relative to the drive amplitude on channel 1. 
This crosstalk is likely caused by the shared first-stage holding capacitors in the demultiplexer and the nonoptimized readout multiplexer circuit. 
Because the current system is intended exclusively for DC voltage output, the effect of such crosstalk can, in principle, be mitigated via waveform-based calibration.

We also evaluate the voltage drop of the holding capacitor.
The holding capacitor connected to channel 1 is first charged to \(+10~\mathrm{V}\).
Then, the switch connected to channel 1 is turned off, and the channel output is monitored for approximately 30~s.
The output voltage is measured using an oscilloscope through a TL031ID buffer amplifier.
After applying a moving average to suppress the periodic noise component, the measured voltage decrease is fitted with a linear function.
Because the observed voltage decrease is minute over the measured time range, reliably extracting an exponential decay time constant from the current data is difficult.
The extracted voltage drop rate at this operating point is \(-19.2(15)~\mathrm{mV/s}\).
For the charging cycle of the 32-channel static-electrode control system, \(1/37.5~\mathrm{kHz} \simeq 26.7~\mu\mathrm{s}\), this drop rate corresponds to a voltage decrease of approximately \(0.51~\mu\mathrm{V}\).
Note that this voltage drop measurement is performed at room temperature, not at 27~K.
Nevertheless, this result provides a reference estimate showing that charge leakage from the holding capacitor is sufficiently slow compared with the relevant TDM update timescale.

Next, the resource requirements of the current 32-channel TDM-based voltage control system are compared with those of a conventional one-to-one electrode control scheme, where each electrode is driven by an independent DAC channel.
In the current TDM system, the room-temperature electronics require only a single DAC channel to generate the multiplexed analog voltage waveform.
The signal lines passing through the vacuum feedthrough comprise one analog voltage line from the DAC and seven logic lines used to control the demultiplexer, totaling eight feedthrough connections.
In contrast, a conventional 32-channel control system requires 32 independent DAC channels and 32 corresponding feedthrough connections to deliver voltages to the electrodes.
Thus, even in the current nonoptimized implementation, the TDM-based approach reduces the number of required feedthrough connections from 32 to 8, corresponding to a fourfold reduction, and the number of required room-temperature DAC channels from 32 to 1.
This comparison highlights the practical advantage of TDM-based voltage control for large-scale ion-trap systems, where the number of electrodes increases rapidly with the number of qubits.

The emphasis of this study is on a proof-of-concept demonstration of TDM-based operation under cryogenic conditions rather than on a detailed characterization of analog signal performance.
Taken together, the results demonstrate the feasibility of TDM-based DC voltage distribution to 32 channels under cryogenic conditions.
Furthermore, the static-electrode demultiplexing system can serve as the downstream 1:M switching stage following the dynamic-electrode demultiplexer output, as illustrated in Fig.~\ref{Fig:archi}(b).


%
\begin{figure}[t]
    \centering
    \includegraphics[width=8.5cm]{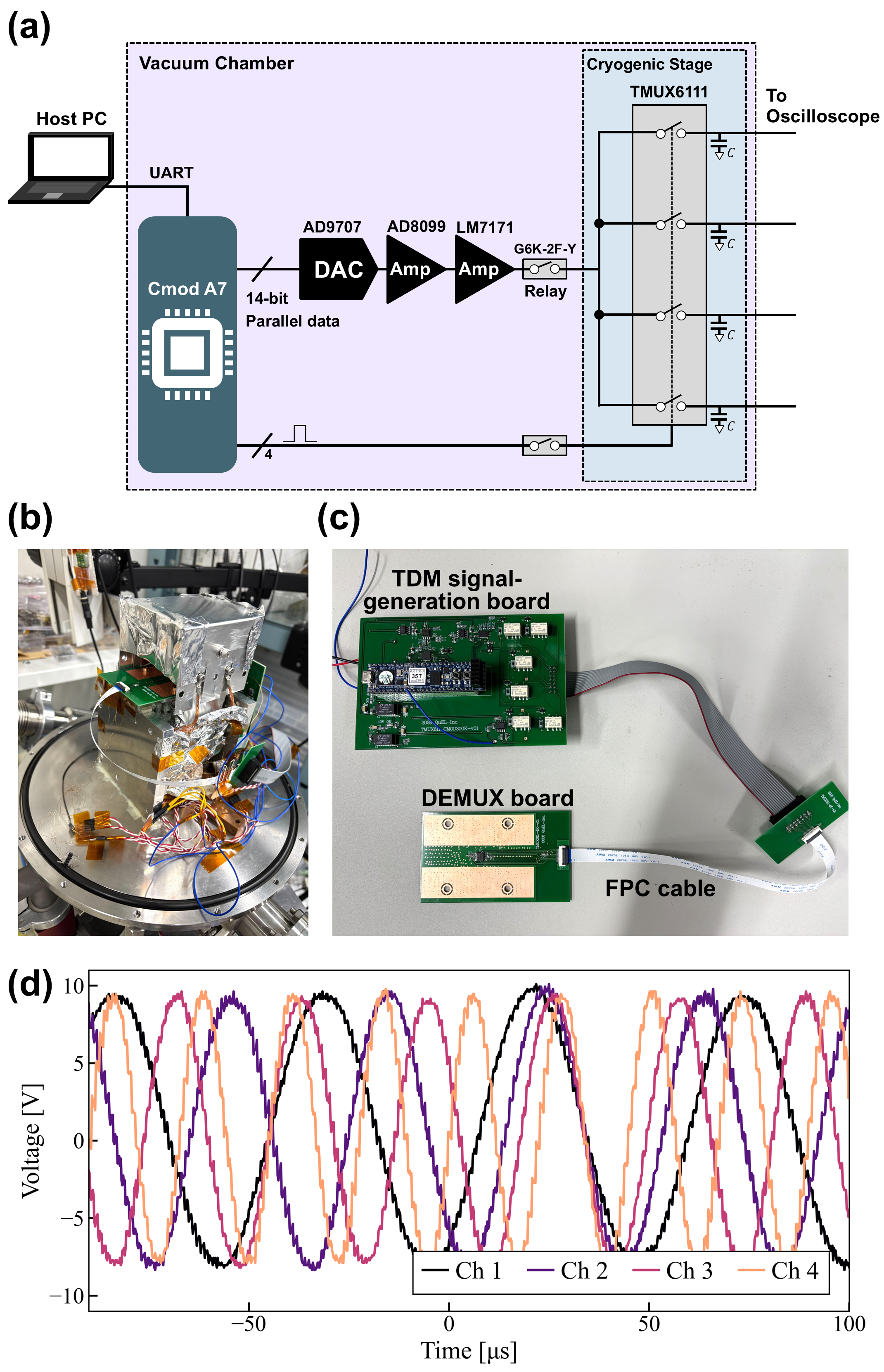}
    \caption{
    TDM-based voltage control system for dynamic electrodes.
    (a) Schematic of system architecture.
    (b) Cryogenic measurement setup with the vacuum chamber removed.
    (c) TDM signal generation board and DEMUX board, connected by a flexible printed circuit (FPC) cable.
    (d) Measured output waveforms from the four-channel demultiplexed system at approximately 14~K.
    }
    \label{Fig:dynamic_control_setup}
\end{figure}

Next, we describe the TDM-based voltage control system developed for dynamic-electrode control and present its cryogenic proof-of-concept evaluation.
We implement a four-channel multiplexed voltage-control system and verify its operation under cryogenic conditions.
A schematic of the overall system is shown in Fig.~\ref{Fig:dynamic_control_setup}(a), and a photograph of the measurement setup is shown in Fig.~\ref{Fig:dynamic_control_setup}(b).

The dynamic-electrode control system has the same overall architecture as the static-electrode control system, comprising a room-temperature TDM signal generation board and a cryogenic DEMUX board.
The TDM signal generation board, shown in the upper part of Fig.~\ref{Fig:dynamic_control_setup}(c), generates time-multiplexed arbitrary waveforms using a DAC (AD9707, Analog Devices, Inc.) driven with 14-bit parallel data from a Digilent Cmod A7 FPGA board.
The FPGA is controlled through the host PC via a universal asynchronous receiver/transmitter (UART) interface.
The DAC operates at an update rate of 4~MHz, and its output is amplified by two cascaded operational amplifiers (AD8099, Analog Devices, Inc.; LM7171, Texas Instruments Inc.).
This corresponds to a per-channel update rate of 1~MHz after four-channel demultiplexing.

The amplified TDM signal is transmitted to the DEMUX board, shown in the lower part of Fig.~\ref{Fig:dynamic_control_setup}(c), which is installed on the cryogenic stage. 
The DEMUX board is implemented using a four-channel switch (TMUX6111, Texas Instruments Inc.) with 120~pF holding capacitors connected to each output.
As shown in Fig.~\ref{Fig:dynamic_control_setup}(c), the output of the TDM signal generation board is routed through a ribbon cable and a conversion connector to the DEMUX board via a flexible printed circuit (FPC) cable. 
The use of a narrow-pitch FPC cable is expected to reduce thermal conduction into the cryogenic stage. 
When cooled using the same GM cryocooler as for the static-electrode DEMUX board, the system reaches a base temperature of 14~K.

To verify the basic operation of the proposed system, we conduct an experiment in which sinusoidal waveforms at four different frequencies are time-multiplexed and subsequently separated by the demultiplexer. 
Figure~\ref{Fig:dynamic_control_setup}(d) shows the measured output waveforms of the four-channel multiplexed system at approximately 14~K. 
The results confirm that the proposed TDM-based architecture successfully delivers each sinusoidal waveform to the corresponding output channel.

Voltage transients with a period of approximately 1~$\mu\mathrm{s}$ are observed at each channel output.
This periodicity is consistent with the 1~MHz per-channel update rate of the system, although the detailed origin of the transients remains unclear.
One possible contribution is interchannel crosstalk.
When identical sinusoidal waveforms are applied to all four channels, the voltage during the hold interval becomes substantially more stable, suggesting that differences between neighboring channel voltages may contribute to the observed transients.
We also evaluate the crosstalk by applying a large-amplitude sinusoidal waveform to channel 1 and measuring the induced signals on the other channels at 14~K. 
The induced amplitudes on channels 2, 3, and 4 correspond to crosstalk levels of approximately -29, -28, and -35~dB, respectively, relative to the drive amplitude on channel 1. 
These results indicate measurable coupling between channels, but do not fully explain the observed transient behavior. 
Therefore, further investigation is required to identify the dominant source of these transients.

Furthermore, the switch used in the current dynamic-electrode control system, TMUX6111, does not include an integrated digital decoder, unlike the TMUX6208 used in the static-electrode control system.
The four switches in the current four-channel demultiplexer are selected using four independent digital control signals.
Although this implementation is sufficient for validating the basic operation of the dynamic TDM-based architecture, future scalable implementations will require a decoder or equivalent logic circuit to reduce the number of digital control lines.


A key finding of this work is that TDM-based voltage control remains functional at cryogenic temperatures. 
Operating the demultiplexer in a cryogenic environment, at approximately 27~K for static-electrode control and approximately 14~K for dynamic-electrode control, is an important step toward reducing the number of vacuum feedthroughs. 
For both static- and dynamic-electrode control systems, the cryogenic operation does not require a logic-control sequence different from that used at room temperature. 
The same demultiplexer control sequence is used at the cryogenic temperature, and no cryogenic-specific adjustment is introduced. 
Under these conditions, the demultiplexers retain their basic switching functionality, and the programmed voltage values are successfully distributed to the individual output channels in both systems. 
These results indicate that the basic demultiplexed sample-and-hold operation is preserved at the cryogenic temperature.
The successful cryogenic validation reported in this study represents a qualitative step forward, confirming that the wiring reduction benefit of TDM can be realized in practice.

The present cryogenic validation is limited to approximately 14~K; operation at few-kelvin temperatures remains to be demonstrated.
Reaching this temperature regime will require cryogenic systems with higher cooling power, improved wiring and thermal anchoring, and further optimization of the overall system configuration.
In addition, quantitative characterization of heat dissipation in the circuit components has not been performed and will be an important step toward system-level thermal budgeting.
A more detailed evaluation of the quality of the analog signal, including output noise and crosstalk, is also required to fully evaluate the suitability of TDM-based voltage control for trapped-ion quantum processors.

In future work, the proposed TDM-based voltage control system will be fully integrated with an ion trap, enabling ion trapping and dynamic control with a substantially reduced feedthrough count. 
The results of this study establish the cryogenic feasibility of this approach and provide a foundation for such integration.


In conclusion, we have developed and cryogenically validated TDM-based voltage control schemes for both principal electrode classes in QCCD trapped-ion systems. 
For static-electrode control, a 32-channel demultiplexed system operates at approximately 27~K, achieving an effective voltage update rate of 37.5~kHz with $\pm10~\mathrm{V}$ per channel. 
For dynamic-electrode control, a four-channel demultiplexed system achieves an effective voltage update rate of 1~MHz at approximately 14~K. 
Together, these results demonstrate that TDM-based voltage control is a viable and scalable solution to the wiring bottleneck in large-scale QCCD systems, enabling independent voltage control of multiple electrodes without a proportional growth in the number of vacuum feedthroughs.

\begin{acknowledgments}
This work was supported by JST (Grant Number JPMJPF2014) and JST Moonshot R\&D (Grant Numbers JPMJMS256G, JPMJMS256F, and JPMJMS256L).
\end{acknowledgments}

\section*{Author Contributions}
\textbf{Ryutaro~Ohira:} 
Conceptualization (equal); 
Data curation (lead); 
Formal analysis (equal); 
Funding acquisition (equal); 
Investigation (equal); 
Methodology (equal); 
Project administration (lead); 
Supervision (lead); 
Visualization (lead); 
Writing - original draft preparation (lead); 
Writing - review \& editing (lead).
\textbf{Shinichi~Morisaka:} 
Conceptualization (equal); 
Formal analysis (equal);
Investigation (equal); 
Methodology (equal); 
Software(lead).
\textbf{Yoshinori~Kurimoto:} 
Formal analysis (equal). 
\textbf{Toshiaki~Inada:} 
Funding acquisition (equal); 
Investigation (equal); 
Resources (equal);
Writing - review \& editing (supporting).
\textbf{Ippei~Nakamura:} 
Conceptualization (equal).
\textbf{Takefumi~Miyoshi:} 
Conceptualization (equal); 
Funding acquisition (equal).
\textbf{Atsushi~Noguchi:} 
Conceptualization (equal); 
Investigation (equal);
Funding acquisition (equal);
Writing - review \& editing (supporting).

\section*{Data Availability}
The data that support the findings of this study are available from the corresponding author upon reasonable request.

\bibliographystyle{aipnum4-2}
\bibliography{ref}

\end{document}